\documentclass[secnumarabic,twocolumn, graphics,floatfix,tightenlines,aps,prb,superscriptaddress]{revtex4-1}

\bibpunct{[}{]}{;}{n}{}{}

 \pdfoutput=1

\usepackage{epsfig}
\usepackage{amsmath,amssymb,amsfonts}
\usepackage{mathrsfs}
\usepackage{bbm}
\usepackage{slashed}
\usepackage{graphicx}
\usepackage{verbatim}
\usepackage[usenames]{color}
\usepackage{mathtools}

\usepackage{stackengine}
\usepackage{hyperref}

\newcommand{\scr}[1]{\mathcal{#1}}

\newcommand{\p}{\partial}

\newcommand{\LCm}{{\scriptscriptstyle -}} 
\newcommand{\LCp}{{\scriptscriptstyle +}}

\newcommand{\LCpm}{{\scriptscriptstyle \pm}}

\newcommand{\ve}[1]{{\bf{#1}}}

\newcommand{\av}[1]{\langle #1 \rangle}

\newcommand{\Tr}{\text{Tr}}

\newcommand{\id}{\text{d}}

\newcommand{\G}{\scr{G}}

\newcommand{\+}{\!+\!}
\newcommand{\m}{\! - \!}

\newcommand{\B}{\mathrm{B}}
\newcommand{\F}{\mathrm{F}}

\newcommand{\SL}{\mathrm{SL}}
\newcommand{\cri}{\mathrm{c}}
\newcommand{\s}{\mathrm{s}}

\begin{document}
	\title{ Suppression of superfluid stiffness near Lifshitz-point instability to finite momentum superconductivity }
	\author{Jonatan W\aa rdh}
	\email[]{jonatan.wardh@physics.gu.se}
	\affiliation{Department of Physics, University of Gothenburg,
		SE-41296 Gothenburg, Sweden}
	\author{Brian M. Andersen}
	\email[]{bma@nbi.ku.dk}
	\affiliation{Niels Bohr Institute, University of Copenhagen, Juliane Maries Vej 30, DK-2100 Copenhagen, Denmark}
	\author{Mats Granath}
	\email[]{mats.granath@physics.gu.se}
	\affiliation{Department of Physics, University of Gothenburg,
		SE-41296 Gothenburg, Sweden}
	
\begin{abstract}
We derive the effective Ginzburg-Landau theory for finite momentum (FFLO/PDW) superconductivity without spin population imbalance from a model with local attraction and repulsive
pair-hopping. We find that the GL free energy  must include up to sixth order
derivatives of the order parameter, providing a unified description of the interdependency of zero and finite momentum  superconductivity. For weak pair-hopping the phase diagram contains 
a line of Lifshitz points where vanishing superfluid stiffness induces a continuous change to a 
long wavelength Fulde-Ferrell (FF) state.   
For larger pair-hopping there is a bicritical
region where the pair-momentum changes discontinuously. Here the FF type state is near degenerate with the Larkin-Ovchinnikov (LO) or Pair-Density-wave (PDW) type state. At the intersection of these two regimes there is a "Super-Lifshitz" point with extra soft fluctuations.  
The instability to finite momentum superconductivity occurs for arbitrarily weak pair-hopping for sufficiently large attraction suggesting that even a small repulsive pair-hopping may be significant in a microscopic model of strongly correlated superconductivity. Several generic features of the model may have bearing on the cuprate superconductors, including the suppression of superfluid stiffness in proximity to a Lifshitz point as well as the existence of subleading FFLO order (or vice versa) in the bicritical regime.

\end{abstract}
\maketitle
\section{Introduction}
Periodically modulated superconductivity is a common theme in several
fields that deal with quantum many-body physics; ranging from cold
atoms and solid-state systems, to dense nuclear
matter\cite{fulde1964superconductivity,larkin1965inhomogeneous,son2006phase,casalbuoni2004inhomogeneous}. Such a state was first considered in systems with a Zeeman split population of spins, referred to as a Fulde-Ferrell-Larkin-Ovchinnikov (FFLO) state. A similar state, but without a symmetry breaking field, is discussed in the context of cuprate superconductors and referred
to as a pair-density wave (PDW)\cite{himeda2002stripe,berg2007dynamical,berg2009theory}. For one, the
PDW state is suggested to account for the suppression of superconductivity in LBCO at $1/8$ doping\cite{moodenbaugh1988superconducting,li2007two}. Moreover, observations in the pseudogap phase, such as a prevalence of diamagnetic response\cite{li2010diamagnetism}, 
arcs in the Fermi surface\cite{baruch2008spectral}, and anomalous quantum oscillations at large magnetic fields\cite{zelli2011mixed,norman2018quantum}, have been put forward as evidence for a more ubiquitous 
PDW state\cite{chakravarty2001hidden,lee2014amperean}. Further, PDW-like states breaking time-reversal symmetry\cite{agterberg2015emergent} have been discussed to account for the apparent finite Kerr-angle\cite{xia2008polar}.
A modulated superconducting state, $\Delta_\ve{Q}$, has been observed in
STM measurements consistent with coexisting superconductivity (SC), $\Delta_0$, and charge-density wave (CDW) $\rho_{2\ve{Q}}$\cite{hamidian2016detection}.
Recently a more direct signature was reported in terms of a double period, $\rho_{\ve{Q}}$, CDW that would follow from coexisting SC and PDW order  \cite{edkins2018magnetic}, thus indicating that the PDW is an intrinsic order in the cuprate superconductors \cite{agterberg2015checkerboard,wang2018pair,dai2018pair}.

Lacking a clear microscopic origin of the PDW
state\cite{fradkin2015colloquium}, effective Ginzburg-Landau (GL)
theories have been utilized to explore the implications of such a
state\cite{berg2009charge,barci2011role,agterberg2008dislocations,agterberg2015emergent,wang2018pair,dai2018pair,boyack2017collective,soto2017higgs,dai2017optical}. 
In this article instead, we start with an effective microscopic model
with repulsive (``$\pi$-phase'')
pair-hopping interactions known to generate PDW states
even in the absence of spin population imbalance\cite{robaszkiewicz1999superconductivity,japaridze2001eta,ptok2009fulde}.
%
\begin{figure}[ht]
	\centering
	\includegraphics[width=0.5\textwidth]{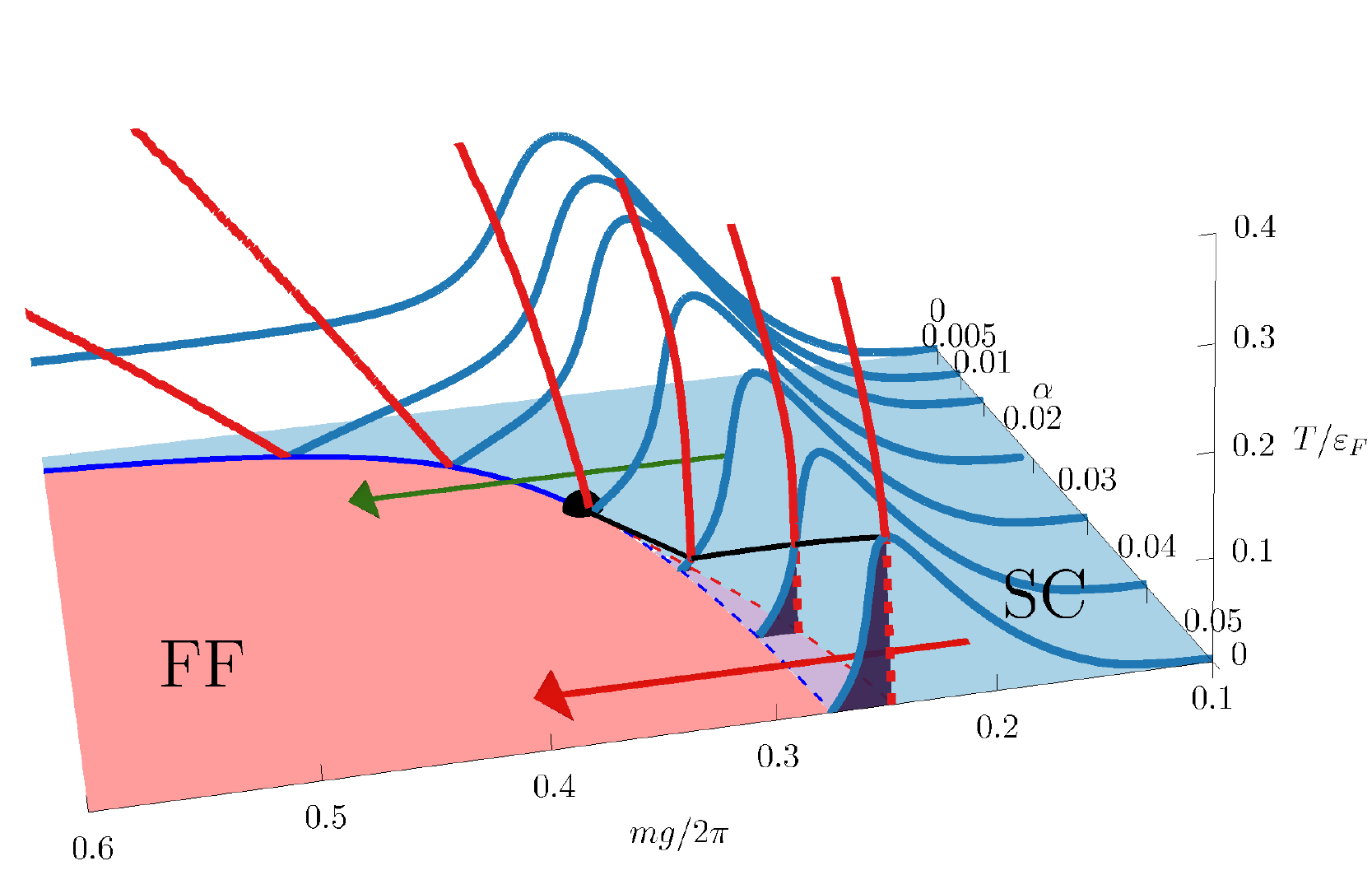} 
	\caption{\label{phaseDiagram} Transition temperatures for
          $Q=0$ superconducting (SC, blue lines) and $Q\neq 0$ Fulde-Ferrell (FF, red lines) states of the model \eqref{Ham},\eqref{Int} as a function of local attraction strength, $g$, and relative pair-hopping strength, $\alpha = g_{\text{pair}}/g_0$. A line of Lifshitz points (given by $(m_{\B})^{-1}=0$ solutions of \eqref{boson_mass}), are shown in solid blue in the $g,\alpha$-plane (dashed for subleading transition). The black solid line indicates a bicritical line. The black dot marks the super-Lifshitz point at the intersection of the bicritical and Lifshitz line. The red and green arrows indicate similar paths as in Figure \ref{picphaseaA}.
	}
\end{figure}
We derive an effective GL theory, that necessarily includes up to sixth order derivatives of the order
parameter, and explore it in the context of the BCS to BEC crossover\cite{nozieres1985bose,randeria1989bound,drechsler1992crossover,de1993crossover,stintzing1997ginzburg}. 
As seen in Figure \ref{phaseDiagram}, the homogeneous state becomes unstable 
to a finite momentum time-reversal breaking Fulde Ferrell (FF)
type state ($\Delta(\ve{r})=\Delta_{\ve{Q}}(\ve{r})e^{i \ve{Q} \cdot
  \ve{r}}$).
This happens at
\textit{arbitrarily} small pair-hopping, $\alpha>0$, for a
sufficiently large attraction, $g$, and occurs through a Lifshitz
point\cite{chaikin1995principles,hornreich1975critical} where the stiffness to deformations of the order parameter
vanishes to lowest order.  For $\alpha>\alpha_{\SL}$ there is
instead a line of bicritical points at finite temperatures, where the
pair-momentum, $\ve{Q}$, jumps, and where the FF state is near degenerate to a translational
symmetry breaking Larkin-Ovchinnikov (LO) PDW state
($\Delta(\ve{r})=\Delta_{\ve{Q}}\cos(\ve{Q} \cdot \ve{r})$). At the intersection of these transitions
$\alpha=\alpha_{\SL}$, $g=g_{\SL}$, $T=0$ there is a special
multicritical ``super-Lifshitz'' (SL) point with extra soft
fluctuations, $\omega\sim q^6$, and distinct mean-field exponents.

Using a unified description of the full momentum dependence of the GL
theory clarifies the interdependence of zero and finite momentum
states. In particular, we emphasize that proximity to
an FFLO state is quite generically expected to suppress the zero momentum
superfluid stiffness, as a result of the GL free energy developing an additional minimum (minima) at finite momentum. Given recent observations of PDW order this may have implications on the general observation of low superfluid stiffness of the cuprate superconductors\cite{PhysRevLett.66.2665,emery1995importance,bovzovic2016dependence}. The formalism also shows that 
superconducting states with dominant uniform order and subdominant PDW
order and states where the roles are reversed, both suggested to exist in the cuprate superconductors\cite{wang2018pair,berg2007dynamical}, are closely related within the same model. 

The paper is organized as follows. In Section \ref{Model} the pair-hopping model is described and the derivation of the Ginzburg-Landau theory is discussed. The mean-field phase diagram in terms of a single momentum dependent theory is considered in Section \ref{GL_section}. Proceeding in Section \ref{BCS_BEC_section} the corresponding BCS to BEC phase diagram (shown in Figure \ref{phaseDiagram}) is derived by including fluctuation effects. In Section \ref{discussion_section} some implications of the model is discussed, such as the suppression of the superfluid stiffness near a Lifshitz point instability in \ref{stiffness}. We conclude with a summary and outlook in Section \ref{outlook_section}.
%
%
%
%
\section{Model \label{Model}} 
 To keep the discussion as general as possible we consider a 2D
 continuous field theory with on-site s-wave pairing and pair-hopping\footnote{In \citet{waardh2017effective} a similar lattice Hamiltonian with nearest neighbor interaction, yielding a d-wave order, was considered.} (setting $\hbar=1,k_{\B}=1$ )
\begin{equation}
\begin{split}
H &=  \int \limits_{\ve{r}}  \psi^\dagger_{\sigma}(\ve{r}) \frac{-\nabla^2 }{2m}  \psi_{\sigma}(\ve{r})  \\
&
-\frac{g_0}{2} \int \limits_{\ve{r}_1,\ve{r}_2 }  T(\ve{r}_{1} \LCm \ve{r}_{2})   \psi^\dagger_{\sigma}(\ve{r}_1) \psi^\dagger_{\sigma'}(\ve{r}_1) \psi_{\sigma'}(\ve{r}_2) \psi_{\sigma}(\ve{r}_2)
\end{split}
\label{Ham}
\end{equation}
%
%
with summation over repeated indices,
and  
%
\begin{equation}
T(\ve{r}_{1} \m \ve{r}_{2}) = \delta( \ve{r}_{1} \m \ve{r}_{2})-\alpha \delta \left( \ve{r}_{1} \m \ve{r}_{2} \pm  \left\{ 
\begin{matrix}
\hat{x} \\
\hat{y}\\
\end{matrix}
\right\} \frac{\lambda}{2} \right) \,. 
\label{Int}
\end{equation}
We define $g_{\text{pair}}=\alpha g_0$ as the strength of the pair-hopping
interaction and 
$g=g_0(1-4\alpha)$ as the strength of zero momentum attraction. We consider $\alpha>0$, i.e. repulsive pair-hopping. 
Further, we consider a type II superconductor and ignore fluctuations
of the gauge field. 

%
%
%
%

We consider the model \eqref{Ham}-\eqref{Int} as an effective model for spontaneous emergence of PDW
superconductivity. However, this sort of pair-hopping interactions have been suggested both as off-diagonal terms of
the microscopic Coulomb interaction \cite{hubbard1963electron,robaszkiewicz1999superconductivity,ptok2009fulde,japaridze2001eta} and as effective interactions in
stripe ordered systems\cite{berg2009theory,waardh2017effective}.
%
%
%
 
\subsection{Method} 
We address the model using a Hubbard-Stratonovich transformation for a bosonic finite momentum pair field $\Delta(\ve{p},i\Omega_m)$ with $\Omega_m= \frac{2m \pi }{\beta}$, a bosonic Matsubara frequency, which couples bi-linearly to the electronic field (see Appendix \ref{Appendix1}). Integrating out the fermions, and expanding to fourth order in the pair field, we write the partition function  $Z=Z_0 \Tr \, e^{-\beta F(\Delta)}$ in terms of the GL free energy functional
\begin{equation}
F = \frac{1}{\beta}  \left( \, \int \limits_{\ve{p}, i \Omega_m}  \Gamma^{\LCm 1}(\ve{p},i \Omega_m) |\Delta(\ve{p},i \Omega_m)|^2 +  \frac{u}{2} \int   |\Delta|^4 \right) 
\label{F0}
\end{equation}
with
\begin{equation}
\begin{split}
&\Gamma^{\LCm 1}(\ve{p},i \Omega_m) =  \\
&\frac{T^{\LCm 1}(\ve{p})}{g_0} - \!\!\!\!
\int \limits_{\ve{k},i\omega_n} \!\!\!\! G(\ve{k} + \frac{\ve{p}}{2},i \omega_n \+ i \Omega_m) G(-\ve{k} \+ \frac{\ve{p}}{2},-i \omega_n) \, 
\end{split}
\label{pair}
\end{equation}
(where the quartic term is written schematically, for details see Appendix \ref{Appendix1}). Here $T(\ve{p})=1-2\alpha \cos(\frac{\lambda}{2} p_x) - 2\alpha \cos(\frac{\lambda}{2} p_y)$ and $G(\ve{k},i \omega_n)=\frac{1}{i \omega_n -\xi(\ve{k})}$, with fermionic Matsubara frequency  $\omega_n = \frac{(2n+1) \pi }{\beta},$ and dispersion $\xi(\ve{k})=\frac{\ve{k}^2}{2m}-\mu$. We expand around the normal state of the dominant mode $\ve{Q}$, $\Delta(\ve{Q}) = 0$,
 thus our theory will hold near $T_{\cri}$. At this mode, the quartic term takes the form 
 \begin{equation}
 u(\ve{Q}) = \int \limits_{\ve{k},i\omega_n} G^2(\ve{k}+\ve{Q},i \omega_n) G^2(-\ve{k},-i\omega_n) \,.
 \end{equation}
 (In the case of expansion around two simultaneous modes additional interactions should also be included as discussed in Section \ref{coexisting}). Further, $\Gamma^{\LCm 1 }(\ve{p},i \Omega_m)$ has a logarithmic UV-divergence which we regularize by introducing a high energy cut-off $\varepsilon_\Lambda= \frac{\Lambda^2}{2m}$ (we set $\varepsilon_\Lambda=80 \varepsilon_{\F}$). 

\begin{figure}[ht]
	\centering
	\includegraphics[width=0.45\textwidth]{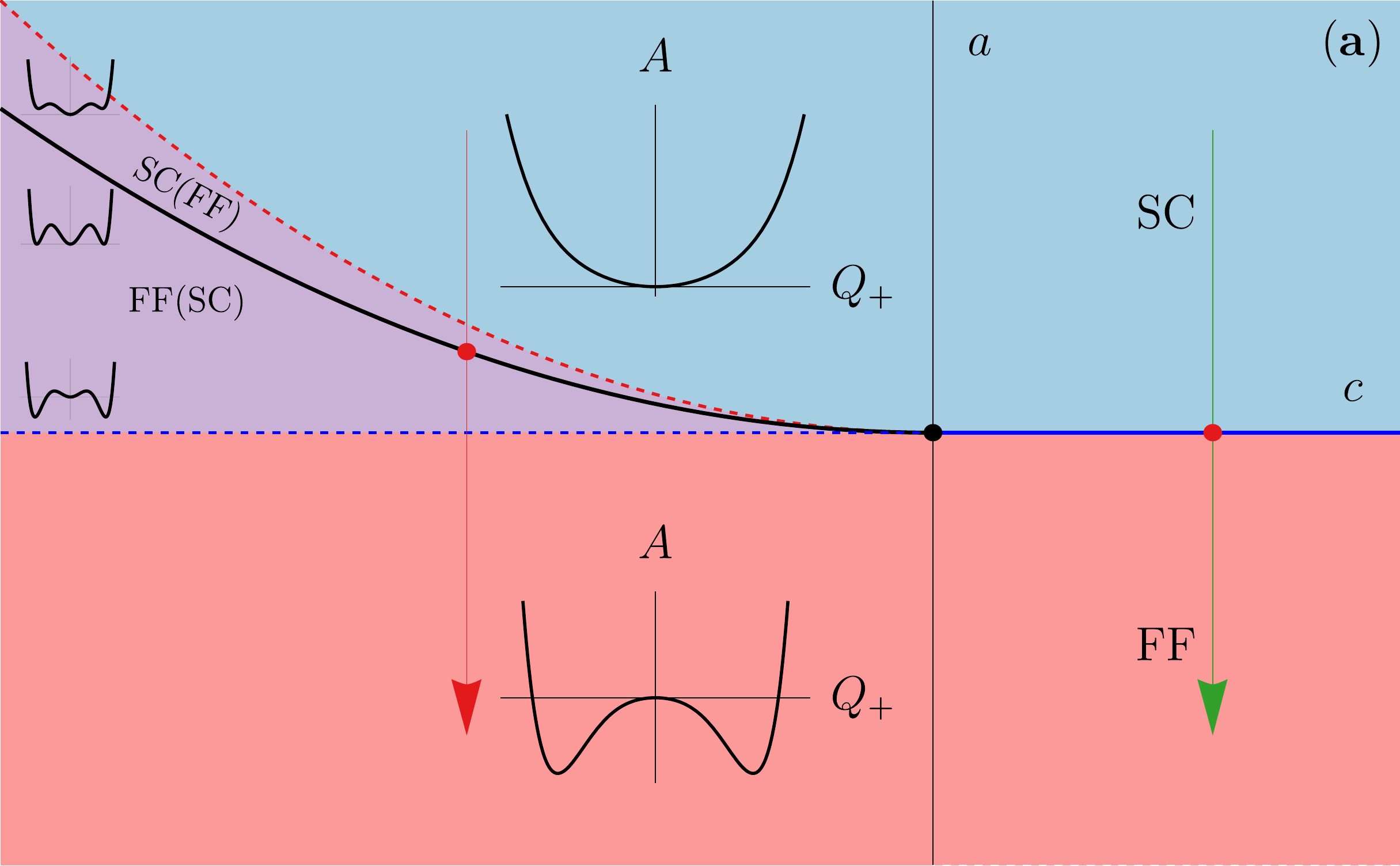} 
	\includegraphics[width=0.45\textwidth]{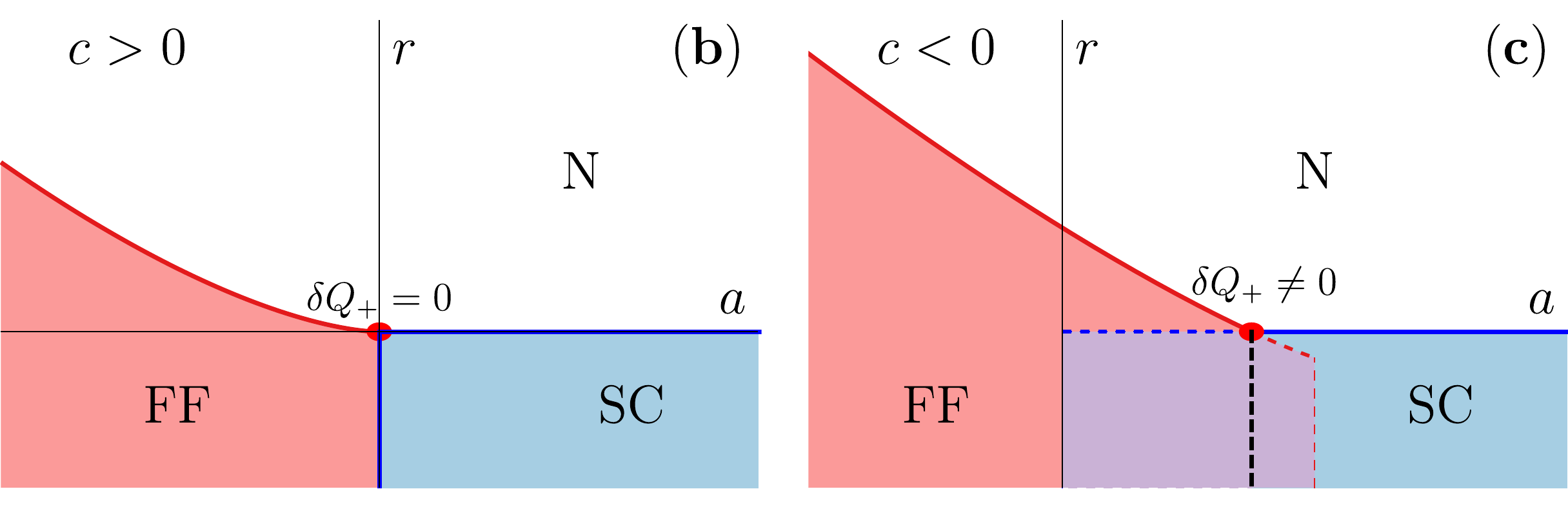} 
	\caption{\label{picphaseaA} Mean-field phase
		diagram of the free energy \eqref{F} (for $Q_\LCm=0$) with SC, $Q_\LCp=0$, and FF, $Q_\LCp \neq 0$, phases. {\bf (a)} Phases as a  function of $a$ ($Q_\LCp^2$) and $c$ ($Q_\LCp^4$), and (inset) the corresponding
		functional form of $A(Q_\LCp)$ for $r=0$. For $c>0$, $a=0$ there is a line
		of Lifshitz points with a continuous transition in
		$Q_+$, as shown in {\bf (b)}. (N.B. when including
		fluctuations the Lifshitz
		point $r=a=0$ is pushed to $T=0$, see Figure \ref{phaseDiagram}). Solid black line $c<0$, $a=\frac{3c^2}{16}$ indicates a line of bicritical points where
		$Q_\LCp$ jumps, as shown in {\bf (c)}. Black
		dashed in {\bf (c) } marks a 1st order transition (vertical assuming $u(\ve{Q})=u$), wheras other dashed lines in  {\bf (a) } and  {\bf (c) }are boundaries of meta-stable subleading
		phases (for $0<a<c^2/4$) that are not thermodynamic transitions. The black dot in {\bf (a)} marks
		the ``super-Lifshitz'' point discussed in the text. The red and green arrows indicate similar paths as in Figure \ref{phaseDiagram}.
	}
\end{figure}
\section{Ginzburg-Landau free energy \label{GL_section}}
To characterize the phase diagram of the model we first study the GL free energy, $F_\text{M}$, which is the static field (mean-field) version of \eqref{F0}. The interaction $T(\ve{p})$ is minimized along the diagonals, thus using rotated coordinates $p_{\LCpm}=\frac{p_x \LCpm p_y}{\sqrt{2}}$ is convenient. We find that for given ($T$, $\mu$, $g$, $\alpha$), $\Gamma^{\LCm 1}(\ve{Q},0)$ can be characterized by a sixth order polynomial in $Q_{\LCp},Q_{\LCm}$:
%
\begin{equation} 
\begin{split}
F_\text{M}&=A(\ve{Q})|\Delta_{\ve{Q}}|^2+\frac{u(\ve{Q})}{2}|\Delta_{\ve{Q}}|^4 \, ,\\
A(\ve{Q})&=r+\frac{a}{2}Q^2+bQ^2_{\LCp} Q^2_{\LCm}+\frac{c}{4}(Q^4_{\LCp}+Q^4_{\LCm})+\frac{1}{6}Q^6 \, . 
\end{split}
\label{F}
\end{equation}
where $a\sim\rho_{\s}$, the superfluid stiffness, as discussed in Section \ref{stiffness}. (For convenience we have fixed the magnitude of the $Q^6$ term.) 

The theory ensures that $u>0$, $b>0$, but $r$, $a$, and $c$ can have
either sign. In minimizing the energy 
 in terms of momentum, $\ve{Q}$, and order parameter, $\Delta_{\ve{Q}}$, we can pick $Q_\LCp$
($Q_\LCm=0$) without loss of generality\footnote{ We anticipate that an explicit band-structure with high density of states directions along $p_x$ and $p_y$ may change the sign of $b$ (as a function of interaction strength) with $b<0$ implying a ``lattice-aligned'' FF state.}. There are in general three possible minima
given by $Q_+=0,Q_+=\pm Q_0$ where $Q^2_0=-\frac{c}{2} +
\sqrt{\frac{c^2}{4}-a}$. The phase diagram in the $(r,a,c)$ space is
outlined in Figure \ref{picphaseaA}. Here we recognize a continuous
evolution of $Q_\LCp$ from a SC to an FF state through a Lifshitz
point (for a vector order parameter\cite{chaikin1995principles}) when
$a$ changes sign and $c>0$. For $c<0$ there is a region of coexisting
local minima at $Q_\LCp=0$ and $Q_\LCp=\pm Q_{0}$. The critical surfaces
meet at a line of bicritical points given by $r=A(Q_{0})=0$ (solid
black line in Figure \ref{picphaseaA}a) where $Q_\LCp$ jumps. That it
is a bicritical transition (1st order), rather than tetracritical
transition (coexisting order), is due to the competition of SC and FF
which we will discuss below.
At the super-Lifshitz point, $a=c=0$,
$A(Q)\sim r+\frac{1}{6}Q^6$ along $Q_{\LCpm}$. The mean-field correlation
length exponent along the soft directions will change from $\nu=1/4$
at a Lifshitz point, to $\nu=1/6$ at the super-Lifshitz
point. Similarly, approaching along $T=T_{\cri}, c=0$ in  the FF state by tuning $a$, the exponent $Q\sim |a|^{\beta_k}$ is
given (in mean-field) by $\beta_k=1/2$ for a Lifshitz point, and $\beta_k=1/4$ for the
SL point\cite{hornreich1975critical}. 
%

\begin{figure}[ht]
	\centering
	\includegraphics[width=0.45\textwidth]{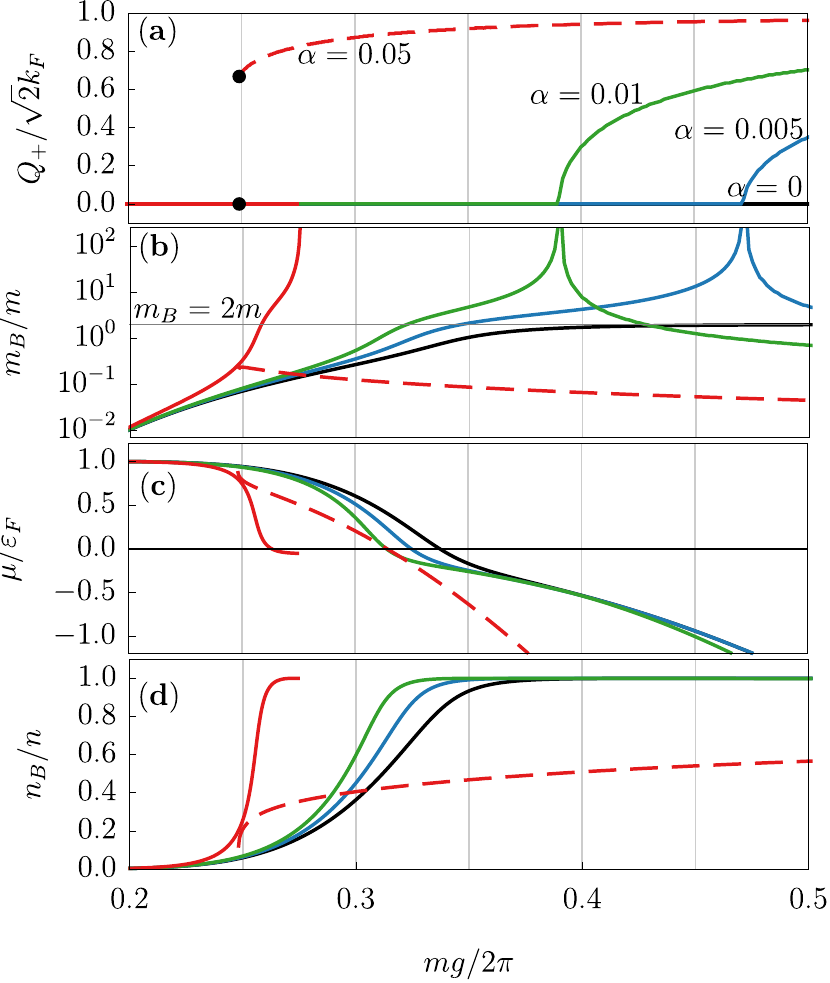} 
	\caption{\label{phase_multi} Evolution of parameters as a function of $g$ for $\alpha=0;0.005;0.01;0.05$. For $\alpha=0.005;0.01$ there is a continuous evolution of $Q_\LCp$ (see {\bf(a)}) starting at the Lifshitz point where the mass $m_{\B}=\sqrt{m_{\B}^{\LCp}m_{\B}^{\LCm}}$ diverges (see {\bf(b)}). For $\alpha=0.05$ there is a jump in $Q_\LCp$, the $Q_\LCp \neq 0$ branch is dashed and the bicritical point is marked with a black dot. For higher $g$ the $\mu$ is pushed to the negative side (see {\bf(c)}) which corresponds to a high occupation of bosons, $n_{\B}$, (see {\bf(d)}).}
\end{figure}
\section{Effect of Gaussian fluctuations \label{BCS_BEC_section}}

To find the phase-diagram within the BCS to BEC crossover we go
beyond mean-field theory by not only considering the 
Thouless criterion
\begin{equation}
 \underset{\ve{p}=\ve{Q}}{\text{min}} \;  \Gamma^{\LCm 1}(\ve{p},0) =0 \, , 
 \label{instability}
\end{equation}
but also the condition for a fixed particle number
\cite{drechsler1992crossover,de1993crossover,stintzing1997ginzburg}  
\begin{equation}
n=\frac{1}{\beta \text{Vol.}}  \frac{\p\ln Z }{\p \mu} = n_\text{F}+n_\text{B} \, , 
\label{particle_number}
\end{equation}
where $n_\text{F}$ is the free fermion density 
and $n_\text{B}$ the contribution from pre-formed pairs (neglecting $\scr{O}(\Delta^4)$). Thus, we solve for $T_{\cri},\mu_{\cri}$ for a given $g,\alpha$. In the weak coupling limit, $g \rightarrow 0$, the pairs are loosely bound with $n_\text{F} \gg n_\text{B}$, yielding the BCS expression $\mu = \varepsilon_\text{F}$, with $\varepsilon_{\F}$ the bare Fermi energy. As the interaction increases, $\mu$ will become negative, with all fermions bound up in pairs ($n_\text{F} \ll n_\text{B}$), which in turn will condense, see Figure \ref{phase_multi}. 

The details of the analysis concerns the general form of $n_{\B} = - \Tr \,  \Gamma \frac{\p \Gamma^{\LCm 1}}{\p \mu}$, which is determined by the analytical structure of the pair-propagator, $\Gamma(\ve{p},z)$, representing the two-particle spectrum\cite{de1993crossover}. In strong coupling $\Gamma(\ve{p},z)$ is well approximated by a simple pole structure yielding (for details see Appendix \ref{Appendix3})
\begin{equation}
n_\mathrm{B} = 2\int \limits_{\ve{q}} \frac{1}{e^{\beta(\frac{r_Q}{\kappa_Q}+\frac{q_i^2}{2m^i_{\B}})}-1} \xrightarrow[\text{reg.}]{\text{3D}} \frac{\zeta(3/2)}{\sqrt{\pi} } T\sqrt{m_{\B}^{\LCp} m_{\B}^{\LCm}}
\label{nb2}
\end{equation}
where we have expanded 
\begin{equation}
\Gamma^{\LCm 1}(\ve{p},z)\approx r_Q-\kappa_Q z+\frac{a_{Q,i}}{2}q_i^2,
\label{a_kappa}
\end{equation}
$i = \pm$ around the saddle point ($\ve{p}=\ve{Q}+\ve{q}$), before performing the Matsubara sum over $z=i\Omega_m$. $m^i_{\B}=\frac{\kappa_{Q}}{a_{Q,i}}$ is the boson mass, reflecting the curvature at the saddle point.
For the $Q=0$ saddle point the mass is isotropic $m_{\B}^\LCpm \propto a^{-1}$ from \eqref{F}; while for the $Q_\LCp>0$ FF state the mass is anisotropic, with $m_{\B}^\LCm \propto (b Q_0^2)^{-1}$ and $m_{\B}^\LCp \propto (-4a-2c Q_0^2)^{-1}$. To regularize the IR-divergence of the 2D bosonic occupation and emulate the  Kosterlitz-Thouless transition \footnote{This regularization captures the dependence of $T_{\cri}$ on the zero temperature phase stiffness, but not the discontinuity of the latter at $T_{\cri}$.}, we follow Stintzing et al. \cite{stintzing1997ginzburg} (see also \citet{gusynin1995toward}) and introduce a third dimension (whose energy-scale equals the thermal expectation value) indicated in \eqref{nb2}.

Note that the lack of a Goldstone mode associated with reorientation of the modulation vector ensures that $n_\text{B}$ remains finite at finite $Q$. This is in contrast to rotational invariant models where $n_\text{B}$ diverges \cite{strinati2018bcs}.

Using \eqref{nb2} will provide a valid description for the strong and weak coupling limit. For intermediate coupling, $\mu \sim 0$ and $n_{\B} \sim n_{\F}$, this analysis could be improved upon by including interaction with scattering states\cite{haussmann1994properties}.

The solution to \eqref{instability} and \eqref{particle_number} was studied for 
a range of values of $\alpha$. The phase diagram is shown in Figure \ref{phaseDiagram} with $Q_\LCp,m_{\B},\mu$ and $n_{\B}$ presented in Figure \ref{phase_multi}. For $\alpha < \alpha_{\SL}$, with $\alpha_{\SL} \approx 0.02$, a transition through a Lifshitz point is realized, and we are moving along a path equivalent to the green arrow in Figure \ref{picphaseaA} as $g$ increases. For $\alpha> \alpha_{\SL}$, $c$ changes sign giving rise to coexisting saddle points along the red arrow as $g$ increases. Note that the qualitative features in Figure \ref{phaseDiagram} and \ref{picphaseaA} can be summarized as the similarity $a \sim g_{\cri} -g$ and $c \sim \alpha_{\SL}- \alpha$.

For small $g$ we see the expected BCS behavior $T_{\cri}\approx1.13\sqrt{
\varepsilon_\Lambda \varepsilon_{\F}}e^{-\frac{2\pi}{m g}}$ (note
$\varepsilon_\Lambda \gg \varepsilon_{\F}$).
We note that $T_{\cri}$ in the FF phase do not saturate due to
continued decrease of effective mass from pair-hopping.
%

\subsection{Weak pair-hopping instability}
%
We now show that any finite $\alpha$
leads to an instability towards FF ($a<0$). First, note that the existence of a Lifshitz point is equivalent to a diverging bosonic mass, or vanishing phase-stiffness (assuming finite $\kappa_Q $), which leads to increased fluctuation, and according to \eqref{nb2} $T_{\cri} \rightarrow 0$, which is clearly seen in Figure \ref{phaseDiagram}. (Strictly speaking \eqref{nb2} is not valid for $m_{\B}=\infty$, and higher order terms in $\ve{q}$ need to be included. However, this does§ not change the conclusion, 
as shown in Appendix \ref{Appendix3}).) $T_{\cri}=0$ ensures that the Lifshitz point will be in the
deep BEC limit, $\mu / T \rightarrow - \infty$. (Because of perfect
nesting  $\Gamma^{\LCm 1}(\ve{p}=0,0)$ can only be non-divergent for $T
\rightarrow 0$ and $g>0$ if $\mu<0$.) The effect of this can be seen in Figure \ref{phase_multi}{\bf (c),(d)}, where increasing $\alpha$ moves the BEC phase to smaller $g$. An exact relation for $m_{\B}$ can be derived in the deep BEC limit 
\begin{equation}
m_{\B}|_{\ve{p}=0} =2m \left(1 - \frac{ \pi  \alpha \Lambda^2\lambda^2  }{mg (1-4 \alpha) \sinh^2 \left(\frac{2 \pi}{mg} \right)}  \right)^{-1} \, .
\label{boson_mass} 
\end{equation}
%
With $\alpha=0$ we find the expected $m_{\B} \rightarrow 2m$. Instead, for $\alpha>0$ there is always a $g=g_{\cri}$ with a divergent mass, proving the existence of the Lifshitz point. $g_{\cri}(\alpha)$ is shown as a blue line in Figure \ref{phaseDiagram}. We also note that this relation is independent of density, i.e. the line of Lifshitz points is stationary with regard to $n$. 
%

For larger $\alpha$, $\alpha> \alpha_{\SL}$ the situation is quite different. Here the zero and finite momentum branches coexist, leading to a bicritical point at which the momentum jumps and almost immediately attains the maximum value $Q_{\LCp,\text{max}}=2 \sqrt{2} / \lambda$. We also see that the bicritical point is on the  weak coupling side with renormalized, yet positive $\mu$. The behavior near the formation of the $Q_\LCp \neq 0$ branch is somewhat intricate (see Appendix \ref{Appendix3}). However, these are meta-stable points without any corresponding thermodynamic transition; thus we marked this part with a dashed line in Figure \ref{phaseDiagram}, extrapolated to $T=0$.

\section{Discussion \label{discussion_section}}
Here we discuss some experimental and theoretical implications of the pair-hopping model for systems without spin-population imbalance.

\subsection{Superfluid stiffness\label{stiffness}}

To discuss the 2D superfluid (or phase) stiffness $\rho_{\s}$ and the corresponding transverse penetration depth $\lambda = \sqrt{d/ 4\mu_0 e^2\rho_{\s}}$ (with $d$ the thickness) we have to make an extrapolation into the ordered state. (As we focus in this work on behavior at $T_{\cri}$, detailed calculations are left for subsequent studies.) Here we will consider only the isotropic case, i.e. the normal $Q=0$ superconductor, where $\rho_{\s}=a_i |\Delta(T)|^2$, and focus on the behavior as a Lifshitz point is approached. Here $a_i$ and $\kappa$ (see below) are as in \eqref{a_kappa}.
%
%
%
As discussed above the fact that a Lifshitz point has $T_{\cri}=0$ ensures that it is in the deep BEC limit where at $T=0$ the pre-formed pairs are all part of the $Q=0$ condensate. Given a proper normalized bosonic field $\psi=\sqrt{\kappa} \Delta$ (with canonical commutation relations) one obtains $\rho_{\s}= \hbar^2|\psi|^2 / m_\B $. At $T=0$ all pairs condense and $|\psi(T=0)|^2=n/2$, with $n$ the full (2D) density of electrons all contributing to the superfluid density, giving $\rho_{\s}(T=0)=\hbar^2 n/2m_{\B}$. For the standard BEC superconductor ($\alpha=0$), where $m_{\B}=2m$ in the BEC limit, this gives the well known result 
$\rho_{\s}(T=0)=\hbar^2 n/4m$, with $m$ the single electron mass. In contrast, non-zero pair-hopping $\alpha>0$ acts as an effective mass term for the pairs according to \eqref{boson_mass}.    
As the mass $m_{\B}$ diverges the superfluid stiffness becomes correspondingly small. In fact, given that $T_{\cri}\propto n/m_{\B}$ (see \eqref{particle_number}) this implies that $\rho_{\s}(T=0)\propto T_{\cri}$ approaching the Lifshitz point. This seems to agree with the general behavior recently observed in the overdoped cuprates\cite{bovzovic2016dependence}, where there is also evidence for broken rotational symmetry close to $T_{\cri}$ \cite{wu2017spontaneous}. However, since our model close to the Lifshitz point is in the BEC limit, it would not be expected to have a Fermi surface even in the normal state (in the some temperature range above $T_{\cri}$) in contrast to overdoped LSCO.  

\subsection{Zero current states}
The minimization of the GL free energy with respect to $\ve{Q}$ is equivalent to the condition that the supercurrent $\ve{J}=dF/d\ve{Q}$ vanishes, as required by Bloch's theorem for ground state currents\cite{bohm1949note}. For an FF state with finite phase velocity the vanishing of the supercurrent is a non-trivial and delicate property (as emphasized already by Fulde and Ferrell\cite{fulde1964superconductivity}) that in the case of a state with population imbalance is satisfied by a corresponding backflow of unpaired quasiparticles. For the model discussed in the present work, even though in the BCS limit there is a backflow due to unpaired quasiparticles (in standard fashion as discussed in Tinkham \cite{tinkham1996introduction}) this is not sufficient to cancel the superflow from the condensate. Instead the cancellation is caused by a backflow of pairs given by the Josephson type supercurrent induced by the pair-hopping \footnote{As discussed in 
\citet{waardh2017effective}, this extra contribution to the current operator arises from the fact that the pair-hopping interaction is not a density-density interaction and consequently does not commute with the polarization operator.}. The existence of an instability to the FF state in the BEC limit (the Lifshitz point) is clearly dependent on this cancellation since there are no unpaired quasiparticles.

A phenomenological GL-model for LO/PDW order\cite{berg2009charge,barci2011role,agterberg2008dislocations,agterberg2015emergent,wang2018pair,dai2018pair,boyack2017collective,soto2017higgs,dai2017optical} consisting of two attractive components at momenta $\pm\ve{Q}$ where both components are locally (meta) stable also implies that there is mechanism for cancelling the supercurrent in the corresponding FF states. Thus, even if the LO/PDW state trivially has zero current due to time reversal symmetry it actually contains at least two Fourier-components that are each expected to satisfy a zero current local stability constraint. This suggests that the elusiveness of PDW order, in for example DMRG studies\cite{dodaro2016intertwined}, for microscopic models with density-density interactions such as the Hubbard or t-J model may in fact related to  limitations in finding zero current solutions with finite phase velocity in such standard models of strongly correlated superconductivity.
\subsection{Coexisting orders\label{coexisting}}
In the FF regime there is a 4-fold degeneracy between states at
$\ve{Q}=\pm Q_{0}\hat{Q}_\LCp$ and $\ve{Q}=\pm Q_{0}\hat{Q}_\LCm$, also,
at the proposed bicritical point there is degeneracy between SC and
FF.
With interactions of the form
$\gamma(\ve{Q},\ve{Q}^\prime)|\Delta_{\ve{Q}}|^2|\Delta_{\ve{Q}^\prime}|^2$
the criterion for coexistence reads
$\gamma(\ve{Q},\ve{Q}^\prime)\leq\sqrt{u(\ve{Q})
  u(\ve{Q}^\prime)}$ with  
 \begin{equation}
 \begin{split}
    &\gamma(\ve{Q},\ve{Q}^\prime)= \\
    &2 \!\!\! \int \limits_{\ve{k},i\omega_n}\!\!\! G(\ve{k}+\ve{Q},i \omega_n) G(\ve{k}+\ve{Q}^\prime,i \omega_n) G^2(-\ve{k},-i\omega_n) \, . 
    \end{split}
 \end{equation}
  We will not discuss this in detail but only infer three important regimes (see Appendix \ref{Appendix2}): (i) For $\ve{Q}\rightarrow 0$ we find $\gamma(0,0)=2u(0)$, which implies that FF is stable for small $\ve{Q}$, i.e. around the Lifshitz point. This also shows that near the super-Lifshitz point the transition is bicritical, as opposed to tetracritical. We have also checked that this holds for larger $\ve{Q}$ near $T_{\cri}$. (ii) At strong coupling, $\mu<0$, the FF state is stable for small enough $Q$, determined by the binding energy of the pairs.
(iii) Then we are only left with the possibility of forming an LO state in weak-coupling, $\mu>0$, for larger $\ve{Q}$. In this case, the FF state depends on parameters in greater details. Extrapolating our model to $T<T_{\cri}$
we find an instability towards the diagonal LO state (PDW) $\Delta_{\ve{Q}}=\Delta_{-\ve{Q}}$, with $\ve{Q}=Q_0 \hat{Q}_\LCp$ (or equivalent). 
Also the LO-FF hybrid type state, $\Delta_{Q_\LCp}=\Delta_{Q_\LCm}$,
such that $\Delta(\vec{r})\sim e^{iQ_xx/\sqrt{2}}\cos (Q_yy/\sqrt{2})$
which breaks both time reversal and translational invariance, is stable
(but subleading to LO) at low temperature. Additional states, e.g. checkerboard
containing all four degenerate FF states are also
possible \cite{soto2014pair,agterberg2008dislocations}. For the LO state one should also consider the interaction $\tilde{\gamma}(0,\ve{Q},-\ve{Q})\Delta_{0}^2\Delta_{\ve{Q}}^*\Delta_{-\ve{Q}}^*+c.c.$ that may turn the 1st order SC-LO transition to a coexistence phase. 

\section{Summary and outlook \label{outlook_section}}
We have seen how an arbitrarily weak repulsive pair-hopping
for sufficiently strong local attraction, $g$, leads to an instability from
zero to finite momentum superconductivity. At weak pair-hopping, this is manifested as a dome of $T_{\cri}$ versus $g$ ending at a superconducting Lifshitz point with a transition into a long-wavelength Fulde-Ferrell state. At larger pair-hopping, the dome is hidden under a bicritical transition where the system changes from SC with subdominant FF or LO order, to a state where the roles are
reversed. At the intersection of Lifshitz and bicritical behavior, there
is a ``super-Lifshitz'' point with extra soft fluctuations and
distinct critical exponents. The Lifshitz transition forces the system
to the strong coupling
regime, $\mu<0$, with pre-formed pairs in the normal state, whereas
the bicritical transition occurs in the weak to intermediate coupling
regime, $\mu>0$. 

The explicit derivation of the GL theory of a finite momentum superconductor from a microscopic model also brings into focus the issue of stabilizing a superconductor with finite phase velocity\cite{fulde1964superconductivity} but zero current, consistent with 
meta-stability of the two ($\pm\ve{Q}$) components of an LO/PDW state. Even for a system with population imbalance, the FFLO order is very delicate\cite{fulde1964superconductivity,SHEEHY20071790,radzihovsky2011fluctuations}, and in the present model with local attraction and without population imbalance it can only be stabilized as an effect of repulsive pair-hopping.

Making connections to the cuprate superconductors it is natural to speculate (given recent evidence for PDW order\cite{edkins2018magnetic}) that the low
$T_{\cri}$ of the underdoped materials may be due to suppressed
phase-stiffness\cite{emery1995importance} caused by proximity to a Lifshitz instability to finite momentum superconductivity. Interestingly, recent observations of low superfliud density in overdoped cuprates\cite{bovzovic2016dependence} also bears a resemblance to the behavior expected approaching such a Lifshitz point. In addition 
there is evidence in the cuprates for both of the regimes with subdominant order discussed in the paper:
(i) Recent evidence for PDW order near vortex cores in BSCCO suggest that
suppression of SC leads to enhancement of PDW order, consistent with subdominant PDW order \cite{agterberg2015checkerboard,edkins2018magnetic,wang2018pair}. (ii) In LBCO, at $1/8$ doping, there is evidence of 2D superconductivity, that has been attributed to interlayer frustrated PDW,
which only at lower temperatures gives way to a 3D Meissner state and
homogeneous SC, consistent with subdominant SC
order\cite{berg2007dynamical,li2007two}.  

Several features of the model remain to be explored further. This includes a more detailed study of the relative prevalence of the various
FF/LO type states, and to include charge order that may additionally favor LO over FF. For this, extending the calculation into the ordered state by self-consistently solving for the pair-field will be necessary. This would also allow for a detailed study of the electromagnetic response of the model with interesting implications such as the anisotropic Meissner state expected in the FF/LO state as well as the interplay between SC and PDW order in the vortex state. The properties of a quantum Lifshitz point\cite{PhysRevB.60.7314} for a superconductor is an interesting topic in its own. 

\section*{Acknowledgements}
Calculations were performed on resources at Chalmers Centre for
Computational Science and Engineering (C3SE) provided by the Swedish
National Infrastructure for Computing (SNIC). The project was
supported by  MAX4ESSFUN. B.~M.~A. acknowledges support from the Independent Research Fund Denmark grant number DFF-6108-00096, and from the Carlsberg Foundation.

\appendix

\section{Effective theory - Hubbard Stratonovich transformation \label{Appendix1}}

Rewriting equation \eqref{Ham}, by going to reciprocal space, and introducing the Nambu-spinor, $ \Psi^\dagger_{\ve{k}} = [ \psi^\dagger_{\uparrow, \ve{k}} \; \psi_{\downarrow, -\ve{k}} ]$, yields
\begin{equation}
\begin{split}
H &=  \int \limits_{\ve{k}}  \Psi^\dagger_{\ve{k}} \, \varepsilon(\tau_3 \ve{k}) \tau_3 \Psi_{\ve{k}}\\ & 
-g_0 \int \limits_{\ve{k},\ve{k}',\ve{p}} T(\ve{p}) \Psi^\dagger_{\ve{k}+\frac{\ve{p}}{2}} \tau_{+} \Psi_{\ve{k}-\frac{\ve{p}}{2}}  \Psi^\dagger_{\ve{k}'-\frac{\ve{p}}{2}} \tau_{-} \Psi_{\ve{k}'+\frac{\ve{p}}{2}} 
\end{split}
\label{Ham2}
\end{equation}
where $\tau_{\pm}=\frac{\tau_1 \pm i \tau_2}{2}$  with $\tau_{i}$ being the Pauli-matrices. Further, $\int \limits_\ve{k} = \int \limits \frac{\id^2 k}{(2 \pi)^2}$, $\varepsilon(\ve{k})=\frac{\ve{k}^2}{2m}$ and 
\begin{equation}
T(\ve{p})=1-2\alpha \cos(\frac{\lambda}{2} p_x) - 2\alpha \cos(\frac{\lambda}{2} p_y) \,. 
\end{equation}
%

The partition function $Z=\Tr e^{-\beta (H- \mu N)}$ can be expressed as a coherent state path integral. We utilize the Hubbard-Stratonovich transformation\cite{fradkin2013field}, which replaces the interaction with a fluctuating bosonic field coupling bilinearly to the electronic field. Thus
\begin{equation}
\begin{split}
&Z= \int \scr{D} \Psi^{*} \scr{D} \Psi \scr{D} \Delta^{*} \scr{D} \Delta  e^{-S(\Psi, \Delta)} \, , \\
& S(\Psi, \Delta) = \int \limits_{k,k'}  \Psi^\dagger_{k} \beta \G^{-1}_{k,k'} \Psi_{k'} + \frac{1}{g_0} \int \limits_{p}  \Delta^{*}(p)T^{-1}(\ve{p})\Delta(p) \, ,\\
&\G^{-1}_{k,k'}=\G_{0,k}^{-1} \delta_{k,k'} -  \Sigma_{k,k'} \,, \quad
\G_{0,k}^{-1}= -i\omega + \tau_3 \xi(\tau_3 \ve{k}) \, , \\
&
\Sigma_{k,k'}=\frac{\Delta(k-k')}{\beta} \tau_{+}  + \frac{\Delta^{*}(-k+k') }{\beta}\tau_{-} \, ,
\end{split}
\label{Znew}
\end{equation}
%
%
where $\xi(\ve{k})=\varepsilon(\ve{k})-\mu$ and $k = (\ve{k},i \omega_n)$, $p=(\ve{p},i \Omega_m)$ for the bosonic and fermionic modes respectively.
%
The total action can be written $S=S^{e}_{\text{eff}}+ S^{\Delta}$ where $S^{\Delta}$ is the second term in \eqref{Znew} and 
\begin{equation}
\begin{split}
&S^{e}_{\text{eff}}(\Delta)=- \Tr \ln \beta \G^{-1} = \\
&- \Tr \ln \beta \G_0^{-1} + \sum_{n} \frac{1}{n} \Tr \left(\G_0  \Sigma \right)^n \, 
\end{split}
\end{equation}
%
%
which is obtained by integrating out the electronic degrees of freedom and expanding around the normal state $\Delta = 0$. Thus, our theory will hold for small $\Delta$, that is, near $T_{\cri}$. Keeping only fourth order terms of $\Delta$ (all odd terms vanish in the Pauli-matrix space) in $S$ we write the partition function as $Z=Z_0 \Tr \, e^{-\beta F(\Delta)}$ where 
\begin{equation}
\begin{split}
&F = \frac{1}{\beta}  \Big( \, \int \limits_{\ve{p}, i \Omega_m}  \Gamma^{\LCm 1}(\ve{p},i \Omega_m) |\Delta(\ve{p},i \Omega_m)|^2 \\
&
+ \frac{1}{2} \!\!\!\! \!\!  \int \limits_{p_1,p_2,p_3} \!\!\!\! \!\! u(p_1,p_2,p_3)\Delta(p_1) \Delta^*(p_2) \Delta(p_3) \Delta^*(p_1\m p_2 \+ p_3)  \Big) \, , \\
&
\Gamma^{\LCm 1}(\ve{p},i \Omega_m) =  \\
&
\frac{T^{-1}(\ve{p})}{g_0} - 
\int \limits_{k}  G(\ve{k} \+ \frac{\ve{p}}{2},i \omega_n \+ i \Omega_m) G(-\ve{k} \+ \frac{\ve{p}}{2},-i \omega_n) \, , \\
&
G(\ve{k},i \omega_n)=\frac{1}{i \omega_n -\xi(\ve{k})} \, .
\end{split}
\end{equation}

\begin{figure}[h!]
\flushleft
	\includegraphics[width=0.48\textwidth]{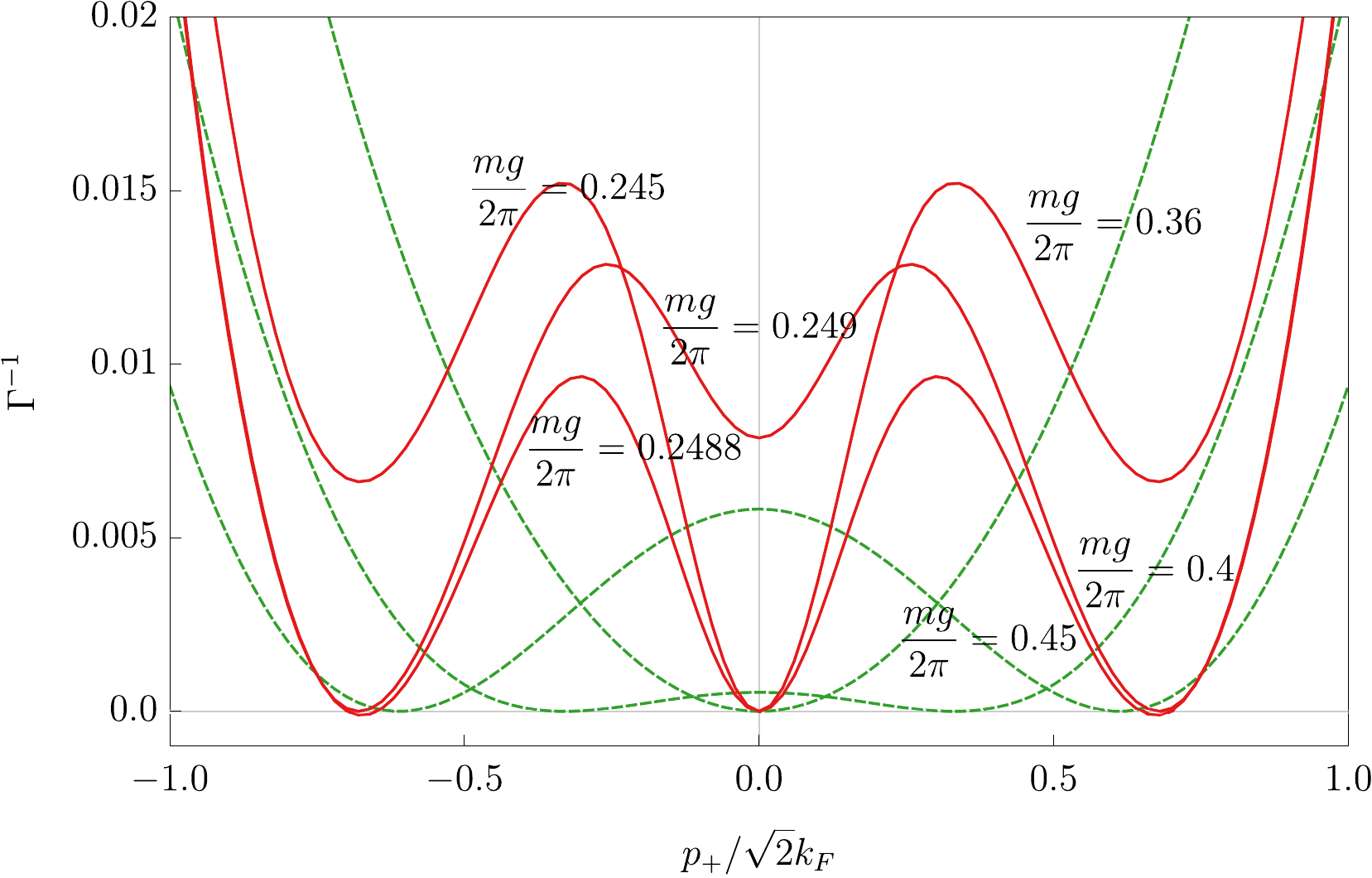} 
	\caption{\label{potential} $\Gamma^{\LCm 1}(p_\LCp,0)$ (measured in $m$) for $\alpha=0.01$(dashed green) and $\alpha=0.05$(solid red) for three different interaction strengths (other variables are the same as in Figure \ref{phase_multi}). For $\alpha=0.01$ we observe a transition through a Lifshitz point where the minimum shifts continuously from $Q=0$ to $|Q_\LCp|>0$ roughly at $\frac{mg}{2\pi}\backsimeq0.4$. For $\alpha=0.05$ we observe a transition through a bicritical point where the momentum changes discontinuously for $\frac{mg}{2\pi}\backsimeq0.2488$. 
	}
\end{figure}

We anticipate that the onset of instability, in general, will occur at finite momenta $\ve{Q}$, i.e.
\begin{equation}
\underset{\ve{p}=\ve{Q}}{\text{min}} \;  \Gamma^{\LCm 1}(\ve{p},0) =0 \, .
\label{instablility_sup}
\end{equation}
In Figure \ref{potential} we show a few realizations of $\Gamma^{\LCm 1}(\ve{p},0)$, as a function of $\ve{p}=p_\LCp \hat{p}_\LCp$ for parameter values presented in Figure \ref{phase_multi}. We clearly see the continuous development of a finite momentum, $p_\LCp = Q_\LCp$, minimum for small $\alpha$, and a discrete jump in momentum for bigger $\alpha$. We also see that the structure of $\Gamma^{\LCm 1}(\ve{p},0)$ is well captured by the characteristic sixth order polynomial presented in equation \eqref{F}, even for large $Q_\LCp$ ($Q_\LCp \sim 2 \pi / \lambda$). We evaluate the repulsive fourth order term at the dominating mode $\ve{Q}$, $u(p_1,p_2,p_3)\rightarrow u(\ve{Q})$ where 
\begin{equation}
u(\ve{Q}) = \int \limits_k  G^2(\ve{k} +\frac{\ve{Q}}{2},i \omega_n) G^2(-\ve{k} +\frac{\ve{Q}}{2},-i\omega_n) \, .
\label{FreeEnergy1}
\end{equation}
%

$\Gamma^{\LCm 1}(\ve{p},i \Omega_m)$ shows a logarithmic UV-divergence which we regularize by introducing a high energy cut-off $\varepsilon_\Lambda= \frac{\Lambda^2}{2m}$. However, in 2D the instability to a superconducting state is in fact equivalent to the existence of a bound state\cite{randeria1989bound,randeria1990superconductivity}. This means that we can express the bare interaction strength, $g$, in terms of the bound state energy in 2D, $E_{\B}$, through the relation $\frac{T^{-1}(0)}{g} = \frac{m}{4 \pi} \ln \left( \frac{2 \varepsilon_{\Lambda}}{E_{\B}} \right)$. Here $\varepsilon_{\Lambda}$ cancels exactly in $\Gamma^{\LCm 1}(\ve{p},i \omega_n)$. In this work we keep the explicit cut-off, however we note that \eqref{instablility_sup} yields $\mu \rightarrow -E_{\B}/2 $, in the strong coupling limit ( $\mu/T \rightarrow - \infty$), i.e. we have to overcome the binding energy in order to break the pair.
%
%

\section{Coexisting orders \label{Appendix2}}
\begin{figure}[h!]
	\includegraphics[width=0.3\textwidth]{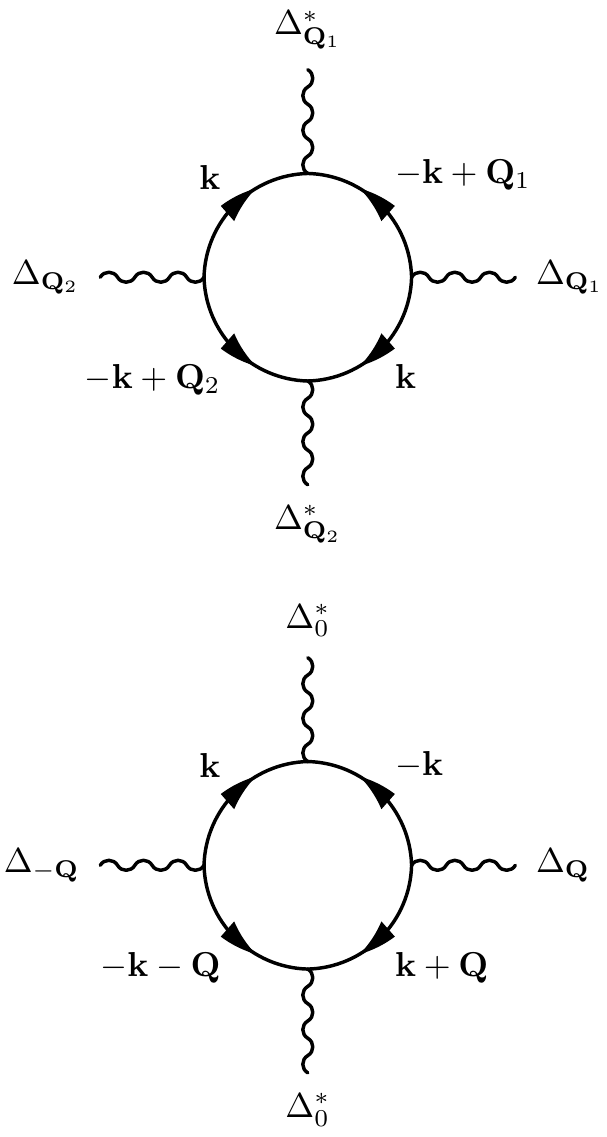} 
	\caption{\label{diag}Diagrammatic representation of \eqref{FreeEnergy2}}
\end{figure}

%
To investigate the possibility of other composite orders we include all anticipated modes, $-\ve{Q},0,\ve{Q},-\overline{\ve{Q}},\overline{\ve{Q}}$ (where $\ve{Q}=Q_{0}\hat{Q}_+$ and $\overline{\ve{Q}}=Q_{0}\hat{Q}_-$), simultaneously. 
We find the following additional terms of the free energy (arising from $u(p_1,p_2,p_3)$)
\begin{equation}
\begin{split}
 &\gamma(0,\ve{Q}) |\Delta_{0}|^2|\Delta_{Q}|^2  \, ,  \gamma(\ve{Q},\ve{Q}') |\Delta_{\ve{Q}}|^2 |\Delta_{\ve{Q}'}|^2, \\
 & \gamma(0,\ve{Q},-\ve{Q}) \Delta_{0}^2\Delta_{\ve{Q}}^{*}\Delta_{-\ve{Q}}^{*} + c.c.
 \end{split}
 \label{additional_terms}
\end{equation}
where
\begin{equation}
\begin{split}
&
\gamma(0,\ve{Q}) = 2 \int \limits_k G^2(\ve{k},i \omega_n) 
G(\ve{k}-\ve{Q},-i\omega_n)G(\ve{k},-i\omega_n) \, , \\
&
\gamma(\ve{Q},\ve{Q}') = 2 \int \limits_k G^2(\ve{k},i \omega_n) 
G(\ve{k}-\ve{Q},-i\omega_n)G(\ve{k}-\ve{Q}',-i\omega_n)  \, , \\
&
\gamma(0,\ve{Q},-\ve{Q}) = \\
&\int \limits_k G(\ve{k},i \omega_n) G(-\ve{k},-i \omega_n) G(\ve{k}+\ve{Q},i\omega_n)G(-\ve{k}-\ve{Q},-i\omega_n)  \, ,
\end{split}
\label{FreeEnergy2}
\end{equation}
represented as Feynman diagrams in Figure \ref{diag}.
(The combinatorial factors arise from cyclic permutation and charge conjugation. However, for the second diagram the charge conjugation is included explicitly in \eqref{additional_terms}.) 

 From \eqref{FreeEnergy1} and \eqref{FreeEnergy2} we see that that $\gamma(0,0)=2u(0)$ and correspondingly $\gamma(\ve{Q},\ve{Q}^\prime)>\sqrt{u(\ve{Q})u(\ve{Q}^\prime)}$ holds for small $\ve{Q}$. This implies that the FF state is stable for small $\ve{Q}$. We can understand this  from the proximity in parameter space to the uniform state. Whereas FF corresponds to a small deformation of the SC, with locally preserved superfluid density, any linear combination of the four FF states will have nodes in the real space pair wave-function and fails to take advantage locally of the full condensation energy.
 
 At strong coupling, $\mu<0$, where the binding energy of the pair, $E_{\B} \sim |\mu|$, is much bigger than the modulation energy, $\sim Q^2$, we anticipate the integrand of $u$ and $\gamma$ to be similar (i.e. for $Q < Q_c$ with $Q_c \propto \sqrt{|\mu|}$) . This was checked numerically and indeed we find $\gamma(\ve{Q},\ve{Q}^\prime)>u(\ve{Q})$, thus FF is stable against forming nodes even in this case.
 
 We are then left with the possibility of forming a LO state in a weak-coupling system, $\mu>0$, at larger $Q$. We do not find instability towards LO at $T_{\cri}$ in the presented parameter regime. But, if we extrapolate below the critical temperature $T<T_{\cri}$ we find an instability towards the diagonal LO state (PDW) $\Delta_{\ve{Q}}=\Delta_{-\ve{Q}}$, with $\ve{Q}=Q_0\hat{Q}_{\LCp}$ or equivalent. However, we are departing from the validity regime of this analysis by extrapolating below $T_{\cri}$. One simple improvement would be to consider expanding around a saddle-point in the ordered phase. This is left for future works.

\section{Bosonic occupation number \label{Appendix3}}
%
%
The contribution to the density of fermions from pre-formed pairs is given by
\begin{equation}
n_{\B} = - \Tr \,  \Gamma \,  \frac{\p \Gamma^{\LCm 1}}{\p \mu} \,  ,
\label{nb}
\end{equation}
which is determined by the analytical structure of the pair-propagator $\Gamma(\ve{p},\Omega)$, describing a two-particle spectrum\cite{de1993crossover}. In 2D there exists a bound state, $\Omega (\ve{p}) =\frac{p_i^2}{2m^{i}_{\B}}$,  at all interaction energies, as well as a two particle continuum represented by a branch cut for $\Omega(\ve{p}) > -2 \mu + \frac{p_i^2}{4m^{i}_{\B}}$. In the weak-coupling limit, where $\mu=\varepsilon_{\F}$, this leads to relaxation dynamics due to decay of weakly bound pairs. Nevertheless, $n_{\B}$ turns out to be negligible because of the high phase stiffness, or small bosonic mass, in accordance with the BCS results. However, in strong coupling limit $\mu \rightarrow -E_{\B}/2 $,
and the branch cut and pole becomes increasingly separated. Thus, the low energy physics is well described by only keeping the, now freely propagating, bound state and \eqref{nb} takes the form
\begin{equation}
\begin{split}
&n_\text{B} = \\
&\int \limits_{\ve{q}}  \left(\exp{\left( \frac{a_{Q,i} q_i^2}{\kappa_Q} \right)} - 1\right)^{-1} \left[- \frac{1}{\kappa_Q}\frac{\p r_Q}{\p \mu} - \frac{\p}{\p \mu} \left( \frac{a_{Q,i} q_i^2}{\kappa_Q} \right)  \right] \, .
\end{split}
\label{nb2_appendix}
\end{equation}
Here we expanded around a saddle point $Q_\LCp=Q_0$ at $\ve{p}=\ve{Q}+\ve{q}$ as $\Gamma^{\LCm 1}(\ve{p},z)\approx r_Q-\kappa_Q z+\frac{a_{Q,i}}{2}q_i^2, i = \pm$ before performing the Matsubara sum over $z=i\Omega_m$. $m^i_{\B}=\frac{\kappa_{Q}}{a_{Q,i}}$ is the boson mass, reflecting the curvature at the saddle point. In the strong coupling limit
 one finds $- \frac{1}{\kappa_Q}\frac{\p r_Q}{\p \mu} =2$ (for $Q_\LCp^2/(2m) \ll |\mu|$). 

At this point we run in to an expected problem, that there is no long range order superconductivity in 2D. This becomes apparent since the Bose-integral in \eqref{nb2_appendix} diverges in 2D. There is however a transition in the Kosterlitz-Thouless (KT) sense, where the low energy state is one with quasi-long range order and a finite super-fluid density. However the physics of the KT-transition is lost by resorting to the Gaussian approximation\cite{stintzing1997ginzburg}. Instead we choose to regularize the divergent integral by allowing the bosons to move out in the third dimension, the $z-$direction. A way to do this is by substituting $\frac{q_i^2}{2m_{\B}^i} \rightarrow \frac{q_i^2}{2m_{\B}^i} + \frac{q_z^2}{2m_{\B}^z}, \int \frac{\id^2 q}{(2\pi)^2} \rightarrow \frac{2 \pi}{\sqrt{\av{q_z^2}}}\int \frac{\id^3 q}{(2\pi)^3}$, where $\av{q_z^2} = 2 m_{\B}^{z} T$ is the thermal expectation value of the momenta in the $z-$direction \cite{stintzing1997ginzburg,wallington2000bcs}. With these considerations we can express the first and second term in \eqref{nb2_appendix} as
\begin{equation}
\begin{split}
&n_{\B}^{(1)} =8 \pi T\sqrt{m_{\B}^{\LCp}m_{\B}^{\LCm}} \int \frac{\id^3 \tilde{q} }{(2\pi)^3} \frac{1}{e^{\tilde{q}^2}-1}  
\, , \\
&
n_{\B}^{(2)} =  \\
&-\frac{16 \pi^2 \alpha \lambda^2}{mg}  T^2  (m_{\B}^{\LCp}+m_{\B}^{\LCm}) \sqrt{m_{\B}^{\LCp}m_{\B}^{\LCm}}\int \frac{\id^3 \tilde{q} }{(2\pi)^3} \frac{\tilde{q}_{\LCp}^2+\tilde{q}_{\LCm}^{2}}{e^{\tilde{q}^2}-1}
\end{split}
\end{equation}
where we introduced the dimensionless momenta $\tilde{q}_i=q_i / \sqrt{2m^{i}_{\B}T}$ and the integration is done over a sphere of infinite radius. The second term, $n_{\B}^{(2)}$, vanishes for $\alpha=0$. But, even for finite $\alpha$, this term can be neglected for small enough densities, $\frac{mg}{2 \pi n_\text{B} \lambda^2} \gg \alpha $ (where $\frac{mg}{2 \pi n_\text{B} \lambda^2} \gtrsim 0.1$, in this work). Thus, we find 
\begin{equation}
n_\text{B}=\frac{\zeta(3/2)}{\sqrt{\pi} } T\sqrt{m_{\B}^{\LCp} m_{\B}^{\LCm}} \, 
\label{nb3}
\end{equation}
presented in \eqref{nb2}. From the above discussion we understand that using \eqref{nb3} will provide a valid description for the strong and weak coupling limit (where it vanishes due to small effective mass). However, at intermediate coupling, when $\mu \sim 0$ and $n_{\B} \sim n_{\F}$, both free and bound fermions coexists and  there is a contribution from scattering states which is not accounted for properly here\cite{haussmann1994properties}. Nevertheless, since the bound state exist for all interaction in 2D (in contrast to the 3D case) there is reason to believe that \eqref{nb3} might still qualitatively give the right description.

\subsection{Corrections to $n_{\B}$}

From the divergence of the bosonic mass we anticipate a suppression of $T_{\cri}$ due to the increase of fluctuation. Indeed, from \eqref{nb3} the divergence of mass is accompanied by a suppression of $T_{\cri}$ to zero. However, strictly speaking, for $m_{\B} \rightarrow \infty $ the derivation of \eqref{nb3} breaks down since the inclusion of higher order kinetic terms necessarily become of importance and will regulate the divergence of $n_{\B}$ as $m_{\B} \rightarrow \infty$ (for fixed $T$). We employ the expansion $\frac{q_i^2}{2m^i_{\B}}+cq^4$, where $m_{\B},c$ are evaluated at the stable point $Q$, and study corrections when $m_{\B}\rightarrow \infty$ for two cases. (Here we neglect cross terms like $bq_{\LCp}^2q_{\LCm}^2$ which do not change the result presented below. Also, note that $b,c$ is not the same as in equation \eqref{F}.)
%
%
%

(i) We start by studying the case when only one mass diverge, say $m_{\B}^{\LCp}\rightarrow \infty$ and $m_{\B}^{\LCm}$ finite. This could happen when the finite $Q$ solution loses its support. We note that the divergence of \eqref{nb2} lies in the IR. Thus, introducing $\tilde{q}_{z,\LCm}=q_{z,\LCm}/\sqrt{2m^{z,\LCm}_{\B}T}$ and $\tilde{q}_{\LCp} =q_{\LCp}^2\sqrt{c/T}$ we can write $n_{\B}$ for small momenta as
\begin{equation}
\begin{split}
&n_{\B}^{(1)} \propto \sqrt{Tm^{-}_{\B}}  \left(\frac{T}{c} \right)^{1/4}\int \tilde{q}^{-1/2} \id \tilde{q}
 \, , \\
&
n_{\B}^{(2)} \propto \sqrt{Tm^{-}_{\B}}  \left(\frac{T}{c} \right)^{3/4} \frac{\alpha \lambda^2}{mg}\int \tilde{q}^{1/2} \id \tilde{q} 
\, .
\\
\end{split}
\end{equation}
(We left out the angular part of the integral, as well as numerical constants.) Both integrals are convergent meaning that $c$ will determine temperature at this point. 

(ii) At the Lifshitz point the mass diverge in two direction, thus for $m_{\B}^{\LCpm}\rightarrow \infty$ we consider
$\tilde{q}_{z}=q_{z}/\sqrt{2m^{z}_{\B}T}$ and $\tilde{q}_{\LCpm}=q_{\LCpm}^2 \sqrt{c/T}$ yielding
\begin{equation}
\begin{split}
&n_{\B}^{(1)} \propto \left(\frac{T}{c} \right)^{1/2}\int \tilde{q}^{-1} \id \tilde{q}
 \, , \quad 
n_{\B}^{(2)} \propto \frac{\alpha\lambda^2}{mg} \left(\frac{T}{c} \right) \int \id \tilde{q} 
\end{split}
\end{equation}
Here, the first term is divergent for finite $T$, thus it forces $T \rightarrow 0$. This means that \eqref{nb3} correctly captures the vanishing of $T_{\cri}$ at the Lifshitz point, this is why we keep \eqref{nb3} even in this case. The second term is finite and negligible also in this case. Nevertheless, $T_{\cri}$ will suffer from corrections near the Lifshitz point.

Further, note that \eqref{nb2} only considers one order at a time. At points of coexistence, like $\frac{mg}{2\pi} \approx 0.25$ for $\alpha=0.05$ where SC and FF are degenerate, one should consider contribution from both orders to $n_{\B}$. However in weak coupling this is not expected to be important since $n_{\F}$ either way dominates. In strong coupling though, it is likely to be important since it redistributes the dominating pre-formed pairs into the different orders.

\subsection{Corrections from time dependent part}
There is an anomalous time-dependent term $\eta_{Q} z q_\LCp$, which arise in the expansion of $\Gamma^{\LCm 1}$ for $Q_0>0$, $\Gamma^{\LCm 1}(\ve{p},z)\approx r_Q-\kappa_Q z+\frac{a_{Q,i}}{2}q_i^2 + \eta_{Q} z q_\LCp, i = \pm$. The inclusion of this term shifts the location of the poles for finite $q_\LCp$ and \eqref{nb2} would take the same form, but with $\kappa_Q \rightarrow \kappa_Q  (1+q_\LCp \eta_{Q}/\kappa_Q)$. We note that, approximately, the highest momenta of relevance in \eqref{nb2} is $q_\text{max}^2 \approx T\kappa_Q/a_{Q}$. From the simulations we have found $q_\text{max} \eta_{Q}/\kappa_Q \lesssim 0.1$ and this term was excluded.

One interesting feature that can be seen from Figure \ref{phase_multi} is that the mass, $m_{\B}=\sqrt{m_{\B}^{\LCp}m_{\B}^{\LCm}}$ attains a finite value at the end of the $Q_0>0$ branch for $\alpha=0.05$. Because of the vanishing curvature, $a_{Q_0,\LCp}=0$, the mass is expected to diverge, since $m_{\B}^{\LCp}=\frac{\kappa_{Q}}{a_{Q,\LCp}}$. However, it turns out that $\kappa_{Q} \rightarrow 0$ in this limit as well, yielding a finite mass. The vanishing of $\kappa_{Q}$ means that we need to consider higher order terms in frequency. This was not done in current work since the ending of this branch corresponds to meta-stable points without any true thermodynamical transition.

\bibliography{Bib_archive}

\end{document}